\documentclass[aps,prl,preprint,superscriptaddress]{revtex4}

\def\ee{\end{eqnarray}}

\def\=:{=\hspace{-.7em}\raisebox{1.1ex}{.}\hspace{.1em}\raisebox{-0.2ex}{.} }

\newcommand {\beq}{\begin{eqnarray}}
\newcommand {\eeq}{\end{eqnarray}}

\newcommand {\1}[1]{\frac{1}{#1}}

\newcommand {\Nf}{N_{\rm F}} 
\newcommand {\Nc}{N_{\rm C}}

\begin{document}

\preprint{TIT/HEP--546}
\preprint{hep-th/0511088}

\title{
Moduli Space of Non-Abelian Vortices
}

\author{Minoru~Eto}
\email[]{meto@th.phys.titech.ac.jp}
\author{Youichi~Isozumi}
\email[]{isozumi@th.phys.titech.ac.jp}
\author{Muneto~Nitta}
\email[]{nitta@th.phys.titech.ac.jp}
\author{Keisuke~Ohashi}
\email[]{keisuke@th.phys.titech.ac.jp}
\author{Norisuke~Sakai}
\email[]{nsakai@th.phys.titech.ac.jp}
\affiliation{ Department of Physics, Tokyo Institute of 
Technology,
Tokyo 152-8551, JAPAN }


\date{\today}

\begin{abstract}
We completely determine the moduli space 
${\cal M}_{N,k}$ of $k$-vortices in 
$U(N)$ gauge theory 
with $N$ Higgs fields 
in the fundamental representation.  
Its open subset for separated vortices
is found as the symmetric product 
$\left({\bf C}\times{\bf C}P^{N-1}\right)^k / \mathfrak{S}_k$. 
Orbifold singularities of this space 
correspond to coincident vortices and 
are resolved 
resulting in a smooth moduli manifold. 
Relation to K\"ahler quotient construction is discussed.

\end{abstract}


\maketitle

\section{Introduction} 
Vortices are very important solitons 
in various area of physics~\cite{Ma}: 
high energy physics, cosmology, 
condensed matter physics and nuclear physics. 
Vortices in Abelian gauge theory have been 
well studied so far \cite{ANO}--\cite{Samols:1991ne}. 
Recently vortices in non-Abelian gauge theory 
(called non-Abelian vortices) have attracted much attention 
\cite{HT}--\cite{NA-vortex} because 
a monopole is confined in the Higgs phase 
with non-Abelian vortices attached 
as a dual picture of quark confinement \cite{vm} 
(see also \cite{vm2,NAvortex2} for related models).
It is very important to determine the moduli space 
of vortices. 
It describes the vortex scattering in $d=2+1$ 
\cite{Samols:1991ne}, is used for 
the reconnection of vortex (cosmic) string in 
$d=3+1$~\cite{Hanany:2005bc,Hashimoto:2005hi}, 
and is needed for the vortex counting in $d=1+1$, 
similarly to 
the instanton counting.  
Identifying vortices with 
certain D-branes 
in a D-brane configuration in string theory, 
the K\"ahler quotient construction of 
the moduli space of non-Abelian vortices 
was suggested \cite{HT}. 
We have determined 
the moduli space of domain walls \cite{walls} 
and other solitons \cite{others} 
by introducing the method of the moduli matrix. 
In this Letter we completely determine the 
moduli space of non-Abelian vortices by applying this 
method.

\section{Vortex equations and their solutions} 
We consider vortex solutions in $d= 3, 4, 5, 6$. 
Field contents are a gauge field $W_M$ ($M=0,\cdots,d-1$), 
two $N \times N$ matrices $H^1$ and $H^2$ of Higgs fields 
and adjoint scalars $\Sigma^I$ ($I=1,\cdots,6-d$). 
The Lagrangian in $d=6$ is
\begin{equation}
{\cal L}_6 
={\rm Tr}\left[-\frac{1}{2g^2} F_{MN}F^{MN}
 + {\cal D}^M H^i ({\cal D}_M H^i)^\dagger 
\right] -V,
  \label{Lagrangian}
\end{equation}
with 
$V=
\frac{g^2}{4}
{\rm Tr}
\bigl[
\left(
H^{1}  H^{1\dagger}  - H^{2} H^{2\dagger} 
- c\mathbf{1}_{N_{\rm C}}
\right)^2 + 4 H^2 H^{1 \dagger} H^1 H^{2\dagger}\bigr]$, 
where the triplet of Fayet-Iliopoulos parameters 
is chosen to the third direction $(0,0,c>0)$. 
This Lagrangian enjoys $U(N)$ gauge symmetry as well as 
$SU(N)$ flavor symmetry. 
By adding fermions this Lagrangian becomes 
supersymmetric with eight supercharges. 
The Lagrangian in $d=3,4,5$
is obtained by trivial dimensional reductions, 
in which the adjoint scalars $\Sigma^I$ 
appear from higher dimensional components 
of the gauge field.  
The scalars $\Sigma^I$  
trivially vanish in vortex solutions 
and we do not need them. 
In either dimension, 
the vacuum is the so-called color-flavor locking phase, 
$H^1 = \sqrt c {\bf 1}_N$ and $H^2=0$  
where symmetry of Lagrangian is broken to 
$SU(N)_{\rm G+F}$. 
This symmetry will be further broken 
in the presence of vortices and therefore 
acts as an isometry on the moduli space.

In the following we simply set $H^2=0$
and $H \equiv H^1$.
The Bogomolnyi completion 
leads to the vortex equations 
\begin{eqnarray}
 0 = {\cal D}_1 H + i {\cal D}_2 H, \quad 
 0 = F_{12} + {g^2 \over 2} (c {\bf 1}_N - H H^\dagger) ,
   \label{BPSeq}
\end{eqnarray}
for vortices in the $x^1$-$x^2$ plane 
and their tension 
\beq
 T = -c \int d^2 x\ {\rm Tr} F_{12} = 2 \pi c k,
  \label{tension}
\eeq
with $k (\in {\bf Z}_{\geq 0})$ 
measuring the winding number 
of the 
$U(1)$ part of broken $U(N)$ gauge symmetry. 

Defining a complex coordinate 
$z \equiv x^1+ix^2$, 
the first vortex equation (\ref{BPSeq}) can be solved as
\begin{eqnarray}
 H =S^{-1} H_0(z),
  \quad
 W_1+iW_2 =-i2S^{-1}{\bar \partial}_z S,   
  \label{solution}
\end{eqnarray}
with $S=S(z,\bar z) \in GL(N,{\bf C})$ 
defined by the second equation (\ref{solution}), 
and $H_0(z)$ an arbitrary $N$ by $N$ matrix 
holomorphic with respect to $z$, which we call 
the {\it moduli matrix}. 
With a gauge invariant quantity 
$\Omega \equiv S S^\dagger$ 
the second vortex equation (\ref{BPSeq})
can be rewritten as
\begin{eqnarray}
 \partial_z (\Omega^{-1} \bar \partial_z \Omega ) 
 = {g^2 \over 4} (c{\bf 1}_N - \Omega^{-1} H_0 H_0^\dagger) .
 \label{master}
\end{eqnarray}
We call this the {\it master equation} for vortices \footnote{
The master equation reduces to the so-called 
Taubes equation \cite{Taubes:1979tm} 
in the $N=1$ case 
by rewriting 
$c \Omega(z,\bar z)=|H_0|^2 e^{-\xi (z,\bar z)}$ 
with $H_0 = \prod_i(z-z_i)$. 
Note that $\log \Omega$ 
is regular everywhere 
while $\xi$ is singular at vortex points.
}. 
This equation is expected 
to give no additional moduli parameters. 
It was proved for the $U(1)$ case \cite{Taubes:1979tm} 
and is consistent with 
the index theorem \cite{HT} in general $N$ 
as seen below.

Eq.~(\ref{master}) implies asymptotic behavior 
$\Omega \to \frac{1}{c} H_0 H_0^\dagger$ for 
$z \to \infty$. Then the tension (\ref{tension})
can be rewritten as 
\beq
 T = 2\pi c k 
 = -i {c\over 2} \oint dz\ \partial {\rm log}({\rm det}H_0) 
 + {\rm c.c.} \label{tension2}
\eeq
We thus obtain the boundary condition on $S^1_{\infty}$ 
for $H_0$ as
${\rm det}(H_0) \sim z^k$ for $z \to \infty$. 
Since any point at infinity $S^1_{\infty}$ must belong 
to the same gauge equivalence class, 
elements in $H_0$ must be polynomial functions of $z$. 
(If exponential factors exist 
they become dominant at boundary $S^1_{\infty}$ 
and the configuration fails to converge to 
the same gauge equivalence class there.) 
From the expression (\ref{tension2}), 
we find that $\det H_0(z)$ has $k$ zeros 
at $z=z_i$ which can be defined as the positions 
of vortices: $\det H_0 (z_i) = 0$. 

There exists a redundancy in the solution (\ref{solution}): 
physical quantities $H$ and $W_{1,2}$ 
are invariant under 
the ``$V$-transformation" 
\beq
 H_0 \to V H_0, \quad S \to V S , 
  \quad \det V = {\rm const.} \neq 0
  \label{V-trans}
\eeq
with $V = V(z) \in GL(N,{\bf C})$, 
whose elements are holomorphic with respect to $z$.  
Here the third condition is  
necessary to maintain the vortex number $k$ unchanged. 
The moduli space ${\cal M}_{N,k}$ for 
$k$-vortices in $U(N)$ gauge theory 
can be formally expressed as a quotient 
\begin{eqnarray} 
{\cal M}_{k,N}={\left\{H_0(z)|H_0(z)\in M_N, 
{\rm deg}\, {\rm det}(H_0(z)) = k \right\}\over 
\left\{V(z)|V(z)\in M_N, 
{\rm det}V(z)={\rm const.}\not=0 \right\}}
\end{eqnarray}
where $M_N$ denotes a set of 
holomorphic $N\times N$ matrices
and ``deg'' denotes a degree of polynomials.

\section{The moduli space of vortices}
The $V$-transformation (\ref{V-trans}) allows us to reduce 
degrees of polynomials in $H_0$ by applying the division 
algorithm.  
After fixing the $V$-transformation completely,
any moduli matrix $H_0$ 
is uniquely transformed to a triangular matrix, 
which we call the standard form, 
\begin{eqnarray}
H_0
 =\left(\begin{array}{ccccc}
  P_1(z)&R_{2,1}(z)&R_{3,1}(z)&\cdots    &R_{N,1}(z) \\
       0& P_2(z)   &R_{3,2}(z)&\cdots    &R_{N,2}(z)\\
  \vdots&          & \ddots   &          & \vdots \\
        &          &          &&R_{N,N-1}(z)\\
       0&\cdots    &          &0         & P_{N}(z)
	   \end{array}\right)
 \label{standard}
\end{eqnarray}
with 
the monic polynomial
$P_r(z) = \prod_{i=1}^{k_r}(z-z_{r,i})$ 
and $R_{r,m}(z)\in {\rm Pol}(z;k_r)$. 
Here ${\rm Pol}(z;n)$ denotes 
a set of polynomial functions 
of order less than $n$. 
The standard form (\ref{standard}) 
has {\it one-to-one correspondence} to 
a point in the moduli space. 
Since ${\rm det}(H_0)=\prod_{r=1}^NP_r(z)\sim z^k$ 
asymptotically 
for $z \to \infty$, 
we obtain the vortex number 
$k = \sum_{r=1}^Nk_r$ from Eq.~(\ref{tension2})
and realize the positions of the $k$-vortices as
the zeros of $P_r(z)$.
Collecting all matrices with given $k$
in the standard form (\ref{standard}) 
we obtain the whole moduli space  ${\cal M}_{N,k}$ for $k$-vortices. 
Its generic points are 
parameterized by the matrix 
with $k_N = k$ and $k_r=0$ for 
$r \neq N$, 
\begin{eqnarray}
H_0 = 
\left(
\begin{array}{cc}
{\bf 1}_{N-1} & - \vec R(z) \\
0 & P(z)
\end{array}
\right)
\label{generic-form}
\end{eqnarray}
where $P(z) = \prod_{i=1}^k(z-z_i)$ 
and $(\vec R(z))^r = R_r(z) \in {\rm Pol}(z;k)$ 
is an $N-1$ vector.
This moduli matrix contains 
the maximal number of the moduli parameters.
The dimension of the moduli space is 
${\rm dim}({\cal M}_{N,k})=2k N$ 
coinciding with the index theorem \cite{HT}. 

The standard form (\ref{standard}) has the merit 
of covering the entire moduli space only once without any 
overlap. 
However, we should parameterize 
the moduli space with overlapping patches 
to clarify the global structure of the moduli space. 
We can parameterize the moduli space  
by a set of ${}_{k+N-1}C_k$ patches defined by 
\begin{eqnarray}
 (H_0)^r{_s} = z^{k_s}\delta^r{_s} - T^r{_s}(z),\quad
 T^r{}_s(z)\in {\rm Pol}(z;k_s). 
\end{eqnarray}
Coefficients of monomials in $T^r{}_s(z)$ are 
moduli parameters as 
coordinates in a patch. 
We denote this patch by 
${\cal U}^{(k_1,k_2,\cdots,k_N)}$.  
We can show that each patch 
fixes the V-transformation (\ref{V-trans}) completely 
including any discrete subgroup, 
and therefore that 
the isomorphism 
${\cal U}^{(k_1,k_2,\cdots,k_N)} \simeq {\bf C}^{kN}$ holds.
The transition functions between these patches 
are given by the $V$-transformation (\ref{V-trans}), 
completely defining the moduli space 
as a smooth manifold,  
${\cal M}_{N,k} \simeq 
\bigcup {\cal U}^{(k_1,k_2,\cdots,k_N)}$.

To see this explicitly we show an example of 
one vortex ($k=1$). 
In this case there exist $N$ patches
\begin{eqnarray}
\!\!\!\!\!
\left(\begin{array}{cccc}
 1&      &0&-b_1^{(N)}    \\
  &\ddots& &\vdots  \\
 0&      &1&-b_{N-1}^{(N)}\\
 0&\dots &0&z-z_0     
\end{array}\right)
\! \simeq \!
 \left(\begin{array}{cccc}
  1&      &-b_1^{(N-1)}  &0 \\
   &\ddots&\vdots & \\
  0&      &z-z_0  &0\\
  0&\dots &-b_{N}^{(N-1)}&1     
 \end{array}\right)\!\simeq \!\cdots
  \label{single}
\end{eqnarray}
Transition functions among these patches are 
given by the $V$-equivalence (\ref{V-trans}) as
$(b_1^{(N)},\cdots,b_{N-1}^{(N)},1)$
$= b_{N-1}^{(N)} 
  (b_1^{(N-1)},\cdots,
   b_{N-2}^{(N-1)},1,b_{N}^{(N-1)})$
$= \cdots$
$= b_1^{(N)} (1,b_{2}^{(1)},\cdots,b_{N-1}^{(1)},b_N^{(1)})$.
These $b$'s are the standard patches for ${\bf C}P^{N-1}$ 
and are called orientational moduli. 
We thus have 
${\cal M}_{N,k=1}$ $\simeq$ ${\bf C} \times {\bf C}P^{N-1}$ 
recovering the result \cite{Auzzi:2003fs}
obtained by 
a symmetry argument.

\section{
Properties of the moduli space
}

We have found that zeros of 
$P_r(z)$ in Eq.~(\ref{standard}) are the positions of 
the vortices. 
We will clarify the meaning of the remaining moduli 
parameters $R_{r,m}(z)$ in Eq.~(\ref{standard}) from now on. 
For simplicity we consider the patch 
${\cal U}^{(0,\cdots,0,k)}$ given in 
Eq.~(\ref{generic-form}) and study $\vec R(z)$ therein.
To this end, we shall introduce basis $\{e^{i}(z)\}\ 
(i=1,2,\cdots,k)$ of the 
space of polynomial ${\rm Pol}(z;k)$.
For example, the simplest complete basis is 
the monomial basis 
$e^i_{\rm m}(z) \equiv z^{i-1}$. Elements of ${\rm Pol}(z;k)$ 
can be expressed by coefficients of monomials in that basis. 
In terms of vortex positions $z_j$ given in the polynomial 
$P(z) = \prod_{i=1}^k(z-z_i)$ 
with degree $k$ in Eq.(\ref{generic-form}), 
we define another basis called {\it point basis} 
(Lagrange interpolation coefficient)
\begin{eqnarray}
 e_{\rm p}^i(z)\equiv 
 \prod_{j=1,(i\not=j)}^k\left({z-z_j\over z_i-z_j}\right),
 \quad 
 e_{\rm p}^i(z_j)=\delta ^i_j .  \label{p-basis}
\end{eqnarray}
The point basis is defined only when $z_i \neq z_j$ for 
$i\neq j$, namely for the separated vortices.
Elements in ${\rm Pol}(z;k)$ can be expressed by values 
at different $k$ points $\{z_i\}$ 
in this basis. For example, $\vec R(z)$ in 
(\ref{generic-form}) can be expressed 
as $\vec R(z) = \sum_{i=1}^k \vec b_i e^i_{\rm p}(z)$ with 
$\vec b_i \equiv \vec R(z_i)$. 
Notice that the $k$ by $k$ matrix $U$ in 
$e_{\rm p}^i(z) = \sum_{n=1}^k {U^i}_n e_{\rm m}^n(z)$ gives 
the Vandermonde determinant 
${\rm det} U^{-1} = \prod_{k\geq j>i\geq1 }(z_j-z_i)$, 
ensuring the completeness of 
the point basis (\ref{p-basis}).   
We thus find 
one-to-one correspondence 
between $\vec{b}_i$ and $\vec R(z)$.

Now we are ready to understand physical meaning of the 
moduli parameters in $\vec R(z)$. To this end,
we consider the infinitesimal $SU(N)$ isometry with 
an element $u(\xi )=\left(\begin{array}{cc}
  {\bf 0}_{N-1}& -\vec \xi  \\ \vec \xi ^\dagger  &0
	   \end{array}\right)$
($\vec\xi$ is an $N-1$ vector)
acting on $H_0$ as
\beq
\delta H_0(z)=v(\xi ,z)H_0(z)+H_0(z)u(\xi ),
\eeq
with an infinitesimal $V$-transformation (\ref{V-trans})
$v(\xi ,z)$ needed to pull-back to 
(\ref{generic-form}). 
This leads to 
\begin{eqnarray}
 \delta \vec R(z) 
   &=&\vec \xi +\vec R(z)(\vec \xi ^\dagger \cdot \vec R(z))
   + \vec s_{\xi ^\dagger }(z) P(z) 
. \label{SU(N)transform}
\end{eqnarray}
Here $\vec s_{\xi ^\dagger }(z)$ 
is a polynomial function for the 
pull-back  
which is uniquely determined for $\vec R$ to be 
in ${\rm Pol}(z;k)$ again.
Noting $P(z_i)=0$ ($i=1,\cdots,k$) 
we obtain $\vec b_i = \vec R(z_i)$ as 
$\delta \vec b_i=\vec \xi +\vec b_i(\vec\xi ^\dagger \cdot \vec b_i)$  
by setting $z=z_i$ in (\ref{SU(N)transform}).
This is precisely the $SU(N)$ transformation law 
for ${\bf C}P^{N-1}$.  
Namely, a set of $(z_i,\vec b_i)$ 
parameterizes ${\bf C}\times {\bf C}P^{N-1}$,
like the moduli of the single vortex mentioned above \footnote{
In the patch ${\cal U}^{(0,\cdots,0,k)}$,
all vortices are aligned in ${\bf C}P^{N-1}$'s   
when $b_i=0$ for all $i$, 
and $b_i$ describe fluctuations from that configuration. 
}.  
Taking into account the fact that
$H_0$ approaches to the one in (\ref{single}) 
for a single vortex 
with the orientational moduli $\vec b_i$  
in the vicinity of 
the $i$-th vortex, 
with $|z-z_i|\ll |z-z_j|$ for all $j (\not=i)$ holding,
we thus find the asymptotic form (open set) of the moduli space 
for separated vortices,  
\begin{eqnarray}
 {\cal M}_{N,k} \leftarrow  
 \left({\bf C}\times{\bf C}P^{N-1}\right)^k / \mathfrak{S}_k
 \equiv S^k \left({\bf C}\times{\bf C}P^{N-1}\right)
 \label{moduli-asympt}
\end{eqnarray} 
with $\mathfrak{S}_k$ permutation group 
exchanging the positions of the vortices \footnote{
This is a half 
of the moduli space 
of $k$ separated $U(N)$ instantons on non-commutative 
${\bf R}^4$, 
$({\bf C}^2 \times T^* {\bf C}P^{N-1})^k/\mathfrak{S}_k$. 
MN thanks Hiraku Nakajima for a comment. 
}. 
Here ``$\leftarrow$" denotes a map to resolve the 
singularities on the right hand side.
Eq.~(\ref{moduli-asympt}) 
can be easily expected from physical intuition; 
for instance the $k=2$ case was found 
in \cite{Hashimoto:2005hi}.
The most important thing is how 
orbifold singularities of 
the right hand side in 
(\ref{moduli-asympt}) are resolved 
by coincident vortices, which we explain below. 
In the $N=1$ case, 
${\cal M}_{N=1,k} \simeq {\bf C}^k/\mathfrak{S}_k$
holds instead of (\ref{moduli-asympt}) \cite{Taubes:1979tm}.

\section{Relation to K\"ahler quotient}
Next we investigate 
the relation between our moduli space and
that from the K\"ahler quotient \cite{HT} 
mainly in the patch ${\cal U}^{(0,\cdots,0,k)}$.
For that purpose, it is important to introduce 
a surjective
map from the space of polynomials ${\rm Pol}(z)$ to 
${\rm Pol}(z;k)$ by
\begin{eqnarray}
q(z) = r(z) + s(z)P(z)
= r(z)\ \ {\rm mod}\ P(z),
\label{mod}
\end{eqnarray}
with $q(z),s(z) \in {\rm Pol}(z)$ and $r(z)\in{\rm Pol}(z;k)$.
The last equality in (\ref{mod}) gives a map from $q(z)$ to $r(z)$
by modulo $P(z)$.
We can extract the moduli parameters 
from $P(z)$ and $\vec R(z)$
as constant matrices ${\bf Z}$ and ${\bf \Psi}$: 
\beq
 z\, e^i(z) &\equiv& 
  ({\bf Z})^i{}_j  e^j(z) \quad 
  {\rm mod~} P(z), \label{mat_z}\\
\left(\begin{array}{c}
       \vec R(z) \\1  \end{array}\right)
&\equiv& ({\bf \Psi})_i e^i(z).\label{mat_psi}
\eeq
When we change the basis as 
$e'^i(z) = U^i{}_j e^j(z)$ 
by $U \in GL(k,{\bf C})$,  
these matrices transform as 
${\bf Z}'=U{\bf Z} U^{-1},\  
{\bf \Psi}'={\bf \Psi } U^{-1}$. 
This is precisely the complexified gauge 
transformation appearing in 
the K\"ahler quotient construction~\cite{HT} 
in which the moduli space is given by $k$ by $k$ 
matrix $Z$ and $N$ by $k$ matrix $\psi$.
The concrete correspondence is obtained by fixing the 
imaginary part of the gauge transformation as 
$
{\cal M}_{N,k} \simeq
\{{\bf Z},{\bf \Psi}\}/\!/GL(k,{\bf C}) 
\simeq 
\left\{(Z,\psi) | \left[Z^\dagger,Z\right]
+\psi^\dagger \psi \propto {\bf 1}_k\right\}/U(k)
$.

For the separated vortices, the point basis (\ref{p-basis}) 
gives us 
${\bf \Psi}$ 
for the orientational moduli and 
the diagonal matrix ${\bf Z}$ whose elements 
correspond to the positions of the vortices  
\begin{eqnarray}
 {\bf Z} = {\rm diag}(z_1,z_2,\cdots,z_k), 
\quad
{\bf \Psi }
 = \left(
 \begin{array}{ccc}
    \vec b_1& \cdots & \vec b_k \\
           1&\cdots  &1 
 \end{array}\right) . \label{matrices}
\end{eqnarray}
As we have mentioned above, 
the point basis (\ref{p-basis}) 
cannot be used for coincident vortices, 
$z_i = z_j$ for $i \neq j$. 
We can deal with them by noting differentiations at $z_i$ 
naturally arise in the limit $z_j \to z_i$.
Let us assume that 
$d_I$ vortices coincide 
at $z=z_I$, 
and divide the labels $i$ to distinguish 
vortices as 
$\{i\} = \{(I,\alpha_I)\}$ with 
$\alpha_I =1,\cdots,d_I$.
We define the {\it generalized point basis} by 
\begin{eqnarray}
&& \qquad e_{\rm p}^{(I,\alpha_I)} (z) \equiv 
 \sum_{n=1}^{k} U^{(I,\alpha_I)}{}_ne_{\rm m}^n(z),  
\label{gen-p-basis} \\
&&\frac{1}{(\alpha_I-1)!}
\frac{d^{\alpha_I-1}e^{(J,\alpha_J)}_{\rm p}(z)}{dz^{\alpha_I-1}}\bigg|_{z=z_I}
= \delta^I{_J}\delta^{\alpha_J}{_{\alpha_I}}
\end{eqnarray}
where $U$ is a $k$ by $k$ invertible matrix, 
whose inverse and determinant are given by
\begin{eqnarray}
&& (U^{-1})^m{}_{(I,\alpha_I)}
={}_{m-1}C_{\alpha_I-1}z_I^{m-\alpha_I} , 
 \\
&& {\rm det} U^{-1} 
 = \prod_{I}\prod_{J>I}(z_J-z_I)^{d_Jd_I},  
\end{eqnarray}
respectively. 
In this basis any function can be expressed by 
a set of differentiations at $z=z_I$.
When no vortices coincide, $d_I=1$ for all $I$, 
the generalized point basis 
(\ref{gen-p-basis}) 
reduces to the point basis (\ref{p-basis}). 
The matrix ${\bf Z}$ 
in the basis (\ref{gen-p-basis}) 
takes the Jordan normal form 
\begin{eqnarray}
 {\bf Z}^{(I,\alpha_I)}{}_{(J,\beta_J)} 
 = \delta^I_J {({\bf z}_I)^{\alpha_I}}_{\beta_J},\; 
 {\bf z}_I=\left(
\begin{array}{cccc}
   z_I&1     &      & 0\\
     0&z_I   &\ddots& \\
\vdots&      &\ddots&1\\
     0&\cdots&0     &z_I
\end{array}\right)  
\end{eqnarray}
and $({\rm \Psi})_{(I,\alpha_I)} = (\Psi_I)_{\alpha_I}$ 
is given by 
\beq
{\bf \Psi}_I =
\left(
\begin{array}{cccc}
\vec R (z_I) & \vec R'(z_I) & \cdots 
    & {1\over (d_I-1) ! }\partial_z^{d_I-1} \vec R (z_I) \\
 1 & 0 & \cdots & 0
\end{array}
\right).
\eeq
Emergence of the Jordan matrix ${\bf Z}$  
is analogous to instantons 
in terms of 
the Hilbert scheme~\cite{Na}.

So far in this section, we have dealt with 
the only patch ${\cal U}^{(0,\cdots,0,k)}$ 
to show correspondence 
between our construction 
and the K\"ahler quotient construction.
In order to complete the correspondence, we have to 
verify it over whole region of the moduli space. 
In what follows we illustrate 
the correspondence in the case of $(N,k)=(2,2)$. 
The moduli space ${\cal M}_{N=2,k=2}$ is 
parameterized by the three patches 
${\cal U}^{(0,2)}$, ${\cal U}^{(1,1)}$, 
${\cal U}^{(2,0)}$ defined in $H_0$'s  
\begin{eqnarray*}
\left(
\begin{array}{cc}
1 & -az-b\\
0 & z^2-\alpha z - \beta
\end{array}
\right) \hspace{-0.1cm}, \hspace{-0.1cm}
\left(
\begin{array}{cc}
z-\phi & -\varphi\\
-\tilde\varphi & z-\tilde\phi
\end{array}
\right) \hspace{-0.1cm}, \hspace{-0.1cm}
\left(
\begin{array}{cc}
z^2 - \alpha z - \beta & 0\\
-a'z- b' & 1
\end{array}
\right) \hspace{-0.1cm},
\end{eqnarray*}
respectively. 
The moduli data in these patches 
can be summarized by  
two matrices ${\bf Z}$ and ${\bf \Psi}$ as follows 
\beq
\left\{{\bf Z},{\bf \Psi}\right\}
= &&\!\!\!\!\!\!
\left\{
\left(
\begin{array}{cc}
     0 & 1\\
 \beta & \alpha
\end{array}
\right),
\left(
\begin{array}{cc}
 b & a\\
 1 & 0
\end{array}
\right)
\right\}, 
\left\{
\left(
\begin{array}{cc}
\phi & \varphi\\
\tilde\varphi & \tilde\phi
\end{array}
\right),
\left(
\begin{array}{cc}
 1 & 0\\
 0 & 1
\end{array}
\right)
\right\},\nonumber\\
&&\!\!\!\!\!\!
\left\{
\left(
\begin{array}{cc}
     0 & 1\\
 \beta & \alpha
\end{array}
\right),
\left(
\begin{array}{cc}
  1 & 0\\
 b' & a'
\end{array}
\right)
\right\}.
 \label{(k,N)=(2,2)}
\eeq
The first one corresponds to the matrices 
$\{{\bf Z},{\bf \Psi}\}$ in Eqs.~(\ref{mat_z}) and (\ref{mat_psi}) 
in the monomial basis.
The $V$-transformation (\ref{V-trans}) between these 
three patches can be expressed 
by the complexified gauge transformation  
between moduli data  
as $\left({\bf Z}',{\bf \Psi}'\right) 
= \left(U{\bf Z}U^{-1},{\bf \Psi}U^{-1}\right)$
with appropriate $U\in GL(2,{\bf C})$.

In conclusion
we have determined the moduli space 
of non-Abelian vortices in 
$U(N)$ gauge theory with $N$ Higgs fields 
in the fundamental representation. 
The orbifold singularity 
appearing in the asymptotic 
form (\ref{moduli-asympt}) 
of separated vortices 
is correctly resolved in the full moduli space,
resulting a complete smooth manifold. 
The relation between our moduli space and the one proposed 
in the D-brane technique is  explicitly shown 
in the case of $N=k=2$. 
The complete identification for general $(N,k)$ 
is an important future work. 
By solving the master equation (\ref{master}) numerically 
we should be able to calculate the moduli metric. 
Refining the discussion of reconnection 
of non-Abelian cosmic string 
\cite{Hashimoto:2005hi} using the moduli metric 
is to be explored. 
We also leave analysis of semi-local vortices 
in $U(\Nc)$ gauge theory with 
$\Nf (>\Nc)$ flavors as a future problem.
Another interesting extension is studying  
non-Abelian vortices on 
Riemann surfaces \cite{Riemann}.


\begin{acknowledgments}
ME, MN and KO (KO) would like to thank Nick Manton 
(Kenichi Konishi) for a useful discussion and are grateful 
to hospitality at DAMTP. 
This work is supported in part by Grant-in-Aid for Scientific 
Research from the Ministry of Education, Culture, Sports, 
Science and Technology, Japan No.13640269 (NS) 
and 16028203 for the priority area ``origin of mass'' 
(NS). 
The work of MN and KO (ME and YI) is 
supported by Japan Society for the Promotion 
of Science under the Post-doctoral (Pre-doctoral) 
Research Program. 
\end{acknowledgments}

\newcommand{\J}[4]{{\sl #1} {\bf #2} (#3) #4}
\newcommand{\andJ}[3]{{\bf #1} (#2) #3}
\newcommand{\AP}{Ann.\ Phys.\ (N.Y.)}
\newcommand{\MPL}{Mod.\ Phys.\ Lett.}
\newcommand{\NP}{Nucl.\ Phys.}
\newcommand{\PL}{Phys.\ Lett.}
\newcommand{\PR}{ Phys.\ Rev.}
\newcommand{\PRL}{Phys.\ Rev.\ Lett.}
\newcommand{\PTP}{Prog.\ Theor.\ Phys.}
\newcommand{\hep}[1]{{\tt hep-th/{#1}}}

\end{document}